# Two-Way Speed of Light and Lorentz-FitzGerald's Contraction in Aether Theory*

Dr. Joseph Levy
4 Square Anatole France, 91250 St Germain lès Corbeil, France
E-mail:levy.joseph@orange.fr

This paper aims at demonstrating that: 1/ Assuming the equality of the two-way transit time of light in vacuo, along the two perpendicular arms of Michelson's interferometers (modern versions of Michelson's experiment), and the anisotropy of the one-way speed of light in the Earth frame, two facts supported today by strong experimental arguments, length contraction (in Lorentz and FitzGerald's approach) should no longer be regarded as an ad hoc hypothesis, it appears necessary and can be easily deduced. 2/ Builder and Prokhovnik had the great merit of having shown that, as a result of to length contraction, the two-way transit time of light along a rod is the same *in all directions in space* (and not only in two privileged directions). We agree with these authors up to this point, but, contrary to what is often believed, their approach failed to reconcile aether theory with the invariance of the *apparent* (measured) two-way speed of light. Yet, as we shall show, due to the systematic measurement distortions entailed by length contraction and clock retardation assumed by aether theory, the two-way speed of light, although anisotropic and dependent on the absolute speed of the frame where it is measured, is always found equal to C. The reasons of this paradoxical but important result will be developed here. They confirm Lorentz-Fitzgerald's contraction and lend support of the existence of a preferred aether frame.

## I. Introduction.

Since the early days of relativity theory, Lorentz-FitzGerald's contraction has been the focus of a debate which is still lively today, and divides physicists in opposite camps.

Some regard length contraction (L.C) as a naïve opinion, for example Wesley [1], Phipps [2], Cornille [3], Galeczki [4]. Some others consider it as a fundamental process which explains a lot of experimental facts. Among them Bell [5], Selleri [6], Builder, Prokhovnik [7], Dishington [8], Mansouri and Sexl [9], Wilhelm [10a].

Length contraction has been proposed by Lorentz and FitzGerald [11] in order to explain the null result of Michelson's experiment. (In fact the result was not completely null, but much smaller than the result expected. We know today that a completely null result can only be observed in vacuum experiments, see later).

Contrary to what is often believed, length contraction is not devoid of physical bases. Not after the year 1901, Joseph Larmor[12] considered a system of two opposite electric charges, describing circular orbits around their common centre of mass. Assuming that the system was in motion through the

---

aether, he demonstrated that the distortion of the electric fields caused by the speed and predicted by classical physics gave rise to the contraction of the system posulated by FitzGerald and Lorentz.

It is worth noting that the conception shared by Lorentz and FitzGerald and by the above mentioned authors about L.C is completely different from the L.C of special relativity. It is the same for all observers. It is not observer dependent. Nevertheless, given that the standard used to measure the length of a moving body is contracted in the same ratio as the body itself, the contraction cannot be demonstrated by an observer attached to the moving frame. Therefore, L.C was never directly observed and only an indirect measurement could be envisaged. This was the objective of different renowned physicists who tried to observe the physical modifications entailed by motion: variation of the refractive index of a refringent solid (Rayleigh [13] and Brace [14]), influence of the aether wind on a charged condenser the plates of which make a certain angle with the direction of translation (Trouton and Noble[15]), experiments of Trouton and Rankine [16] and of Chase [17] and Tomashek [18] on the electrical resistance of moving objects, and finally of Wood Tomlison and Essex [19] on the frequency of the longitudinal vibration of a rod.

Yet the experiments proved all negative.

However the lack of experimental evidence, could be explained by the fact that length contraction was veiled by the increase of mass with speed following the law $m = m_0 \gamma$ (see ref [20]), an explanation to which Lorentz had recourse. Yet in the light of the recent developments of physics, this argument is not without raising some objections which need to be responded accurately. Indeed it is necessary to prove that the experimental law $m = m_0 \gamma$ is not in contradiction with the Lorentz aether concept. The question has been already studied in earlier publications[21-23]. We develop our more recent approach of the subject and give a response in the appendix.

A more recent experiment by Sherwyn[24] also yielded a negative result. The author considered an elastic rod of length L rotating about one of its ends in the laboratory frame. At low rotation rates, the length of the rod adiabatically follows the value demanded by the equilibrium lengths of the molecular bonds. Obviously, this length cannot be estimated by laboratory meter sticks, since they show the same dependence of length on angle. However, according to the author, at high rotation rates, when the time required to rotate 90° becomes comparable to the period of vibration of the structure, the macroscopic length would not be able to exactly follow the "bond equilibrium length." a fact which should make it possible to highlight length contraction.

To support his demonstration, the author assumed that "*the relativistic contraction is a physical process and proceeds with the speed of sound in the structure*" and "*it will occur relatively slowly in a time comparable to L/v where v is the speed of sound in the rod*".

This statement is not based on experimental grounds and nothing proves that it corresponds to reality. In his book "Light in Einstein's universe", Prokhovnik[7] objects that the contraction should occur in a time comparable to L/C. In any cases there is no certainty that under the conditions of the experiment the adiabatic process would not have occurred..

We must add that for an aether drift estimated at 300Km/sec the variation of L due to length contraction would have been of the order of $\frac{1}{2} 10^{-6}$ L, which for a spring of 1 metre long would yield a contraction of 1/2000 mm or half a micron, a length very difficult to highlight. Yet the spring used in Sherwyn's experiment measured 0.123 m.

For these reasons Sherwyn's experiment proves unconvincing.

Another argument which questions Sherwyn's experiment has also been given by D. Larson[25].

Note that, at first sight, we can also object to Lorentz-Fitzgerald's contraction that the compressibility of matter is limited, and length contraction seems difficult to justify at very high speeds. For example at 0.9999 C the ratio $L/L_0$ would be reduced to 1.4%.

But we can answer that the law has been proposed following an experiment performed at low speed (Michelson's experiment). It would not adopt exactly the same form at very high speeds.

Today, as we shall see, strong arguments exist in support of Lorentz-Fitzgerald's contraction (L.C.). One of these arguments is that L.C enables to explain (in all directions of space and not only in two perpendicular directions) the isotropy of the *apparent* (measured) two way speed of light. L C sets up the aether assumption on solid bases which enables it to be confronted with special relativity. (Of course it is a question here of Lorentz's aether theory based on the existence of a preferred aether frame in which the aether is at rest).

## II. Lorentz-FitzGerald's contraction explains the null result observed in *vacuum* Michelson's experiments.

We know today that, even if we use two clocks to make the measurement, the standard synchronization procedures (Einstein-Poincaré procedure or slow clock transport) only allow the measurement of the two way speed of light [7,26].

As we shall show, the aether theory maintains that the value C is found because the two-way speed of light is measured with contracted meter sticks and clocks slowed down by motion.

According to Anderson, Vetharaniam and Stedman [27], all the recent experiments purporting to illuminate the isotropy of the one way speed of light were based on erroneous ideas (because they considered that the slow clock transport procedure allows exact synchronization).

On the contrary, a number of arguments are put forward today in favour of the anisotropy of the one-way speed of light, when no distortion alter the measurement. Although its direct estimate comes up against major difficulties, several authors applied themselves to evaluate it from the measurement of the terrestrial aether velocity, based on the fact that light signals propagate isotropically in the aether frame. (Note that the orbital velocity of the Earth is of the order of 30Km/sec. Therefore, as a first approximation, the *absolute speed* of the Earth can be identified with the solar system absolute velocity which, as we shall see, is estimated at about 330-400 Km/sec).

A first evaluation of the solar system absolute velocity was already made in 1968 by De Vaucouleurs and Peters, who measured the anisotropy of the red shift relative to many distant galaxies. The experiment was made again by Rubin in 1976.

A more reliable estimate was obtained by measuring the anisotropy of the 2.7° K microwave background radiation, uniformly distributed throughout the Universe. "An observer moving with velocity *v* relative to the microwave background can detect a larger microwave flux in the forward direction (+*v*) and a smaller microwave flux in the rearward direction (-*v*). He can observe a violet shift in the forward direction (+*v*) and a red shift in the rearward direction (-*v*) (Wilhelm)".

From this data, the absolute velocity of the solar system could be measured ((Conklin (1969), Henry (1971), Smoot et al (1977), Gorenstein and Smoot (1981), Partridge (1988)). Let us also quote the method of measurement based on the determination of the muon flux anisotropy (Monstein and Wesley (1996)). An assessment of all these experiments is given by Wilhelm [10a] and Wesley [10b].

Another verification of the absolute speed of the Earth frame was made by Roland De Witte in 1991. To this end, 5 Mhz radio frequency signals were sent in two opposite directions through two buried co-axial cables linking two caesium beam clocks separated by 1.5 km. Changes in propagation times were observed and were recorded over 178 days. De Witte interpreted the results as evidence of absolute motion. (Unpublished, cited by Cahill).

More recently (april 2003), Cahill and Kitto reinterpreted the Michelson and Morley experiment. They asserted that Michelson interferometers operating in gas mode are capable of revealing absolute motion[28a]. They analysed the old results from gas-mode Michelson interferometers experiments which always showed small but significant effects. The authors asserted that after correcting for the air, the Miller experiment gives an absolute speed of the Earth frame of v=335±57 Km/sec. A more recent assessment by Cahill[28b] yields a value of about 400Km/sec.

Marinov[29] also attested having measured the absolute velocity of the solar system by means of different devices (coupled mirrors experiment, toothed wheels experiment). The experiments are described in detail in the book of Wesley [1], and are quoted by Wilhelm [10a].

According to Wesley[1,10b,30] the Marinov (1974, 1977a, 1980b) coupled mirrors experiment is one of the most brilliant and ingenious experiment of all time. It measures the very small quantity v/C where v is the absolute velocity of the observer by using very clever stratagems.

On the basis of his experiments, the author asserted that the absolute velocity of the solar system is of the order of 320±20 Km/sec.

This result was in agreement with most of the observations and experiments described above which lend support to the existence of a fundamental inertial frame, whose absolute speed is zero, but whose relative speed with respect to the Earth frame is of the order of 330-400 km/sec.

Consider now a Michelson interferometer whose longitudinal arm is aligned along the $x_0$-axis of a coordinate system $S_0$ (0, $x_0$, $y_0$, $z_0$) attached to the Cosmic Substratum. The device is at rest in the Earth frame which, during the short time of the experiment, is assumed to move along the $x_0$-axis at speed v.

It is easy to verify that, in reply to the statement that the speed of light is C - v in the + $x_0$ direction, and C + v in the opposite direction, the arm will be contracted in the ratio:

$$\ell = \ell_0 \sqrt{1 - v^2/C^2} \qquad (1)$$

where $\ell$ is the length of the arm in the Earth frame, and $\ell_0$ its length when it is at rest in the aether frame.

With the same starting point, we shall show that the *apparent* (measured) two-way speed of light along the $x_0$-axis, is found equal to C independently of the speed v. Let us demonstrate formula (1).

A priori, we do not know if $\ell = \ell_0$ or not. The two-way transit time of light along the longitudinal arm will be:

$$t_1 = \frac{\ell}{C-v} + \frac{\ell}{C+v} = \frac{2\ell}{C(1 - v^2/C^2)}. \qquad (2)$$

Now, in the arm perpendicular to the direction of motion, there is no length contraction.

According to aether theory, the speed of light is C exclusively in the aether frame. The signal starts from a point P in this frame towards a point O at the end of the arm and then comes back towards point P'. During that time, the interferometer has covered the path $vt_2$ (see figure 1).

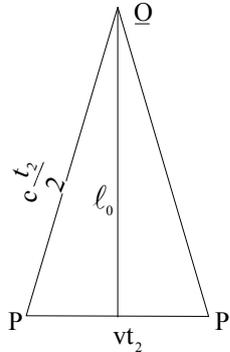

Figure 1.

The path of the light signal along the arm perpendicular to the direction of motion viewed by an observer from the aether frame.

we have

$$\left(C\frac{t_2}{2}\right)^2 - \left(v\frac{t_2}{2}\right)^2 = \ell_0^{\,2}$$

$\Rightarrow \qquad \ell_0 = \frac{t_2}{2}\sqrt{C^2 - v^2}$

so that

$$t_2 = \frac{2\ell_0}{C\sqrt{1 - \frac{v^2}{C^2}}}. \qquad (3)$$

*In vacuum experiments* the displacement of the fringes when we change the orientation of the interferometer is hardly perceptible. Neglecting this fringe shift[*] which is really too small to explain the existence of an aether drift of about 400 km/sec, we can write $t_1 = t_2$, that is:

---

[*] Note that, the modern versions of Michelson's experiment in vacuo greatly confirm the equality of the two-way transit time of light along the two arms of the interferometer. The measurements made by Joos (1930), Jaseja et al (1964), Brillet and Hall (1971) have verified this result which was almost perfect for Brillet and Hall. For a review of the topic consult H.C Hayden, Phys essays 4, 36, (1991). More recent confirmations have been made by Müller et al, 28 May 2003, Sherrmann et al, 15 Aug 2005 and Shiller et al, 18 Oct 2005.

$$\frac{2\ell}{C(1-v^2/C^2)} = \frac{2\ell_0}{C\sqrt{1-v^2/C^2}}.$$

Hence
$$\ell = \ell_0\sqrt{1-v^2/C^2}.$$

Therefore, if we take the anisotropy of the one way speed of light into account, length contraction must no longer be considered as an ad hoc hypothesis. On the contrary, it must be seen as a necessary cause of the Michelson result.

Now, on account of clock retardation, the *apparent* (measured) two-way transit time of light will be (from (3)):

$$\frac{2\ell_0}{C}.$$

In aether theory, length contraction is a real process valid for all observers. It is not observer dependent. Since the length of the longitudinal arm is determined with a standard contracted in the same ratio as the arm, it is found equal to $\ell_0$ and not to $\ell$, so that the *apparent* (measured) two way speed of light along the $x_0$-axis will be found equal to C. (It is in fact different from its real value, which according to formula (2) is $C(1-v^2/C^2)$.

NB - In the absence of length contraction, the *apparent* two-way speed of light would not have been found equal to C, in contradiction with the experiment.

## III. Lorentz-FitzGerald's contraction explains the *apparent* speed of light invariance.

- But this is not all. We will now demonstrate that L.C explains the independence of the *apparent* two way speed of light from any direction of space and from the absolute speed v of the 'inertial' system where it is measured..

The demonstration is based on Builder and Prokhovnik's [7] studies whose importance is indisputable but, as we shall see, some of the conclusions of Prokhovnik were questionable and could not enable to demonstrate that this *apparent* velocity is C (which is the real speed of light in the aether frame).

Consider two co-ordinate systems, $S_0$ is at rest in the cosmic substratum, and S is attached to a body which moves with rectilinear uniform motion along the $x_0$-axis of the $S_0$ system and suppose that a rod AB making an angle $\theta$ with the $x_0$, x-axis, is at rest with respect to the system S (see figure 2).

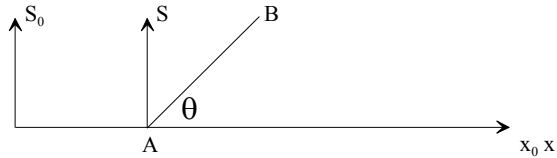

Figure 2

The rod AB is at rest with respect to frame S

At the two ends of the rod, let us place two mirrors facing one another by their reflecting surface, which is perpendicular to the axis of the rod $\ell = AB$. At the initial instant, the two systems $S_0$ and S are coincident. At this very instant a light signal starts from the common origin and travels along the rod towards point B. After reflection the signal returns to point A.

We do not suppose a priori that $\ell = \ell_0$ (where $\ell_0$ is the length of the rod when it is at rest in the aether system $S_0$). We remark that the path of the light signal along the rod is related to the speed $C_1$ by the relation:
$$C_1 = \frac{AB}{t} \qquad \text{(see figure 3)}$$
where t is the time needed by the signal to cover the distance AB.

In addition, when the signal reaches point B, the system S has moved away from $S_0$ a distance:
$$AA' = vt$$

so that: $$v = \frac{AA'}{t}.$$

Now, from the point of view of an observer at rest in $S_0$, the signal goes from point A to point B' (see figure 3)

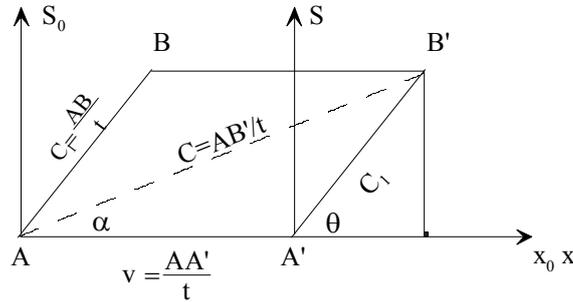

Figure 3

The speed of light is equal to C from A to B', and to $C_1$ from A' to B'.

C being the speed of light in $S_0$, we have:
$$\frac{AB'}{t} = C$$

and hence, the projection along the x-axis of the speed of light $C_1$ relative to the system S, will be equal to $(C \cos \alpha - v)$.

We remark that: $C \cos \alpha - v = C_1 \cos \theta$.

The three speeds, C, $C_1$ and v being proportional to the three lengths AB', AB and AA' with the same coefficient of proportionality, we have
$$C^2 = (C_1 \cos \theta + v)^2 + C_1^2 \sin^2 \theta$$
Therefore: $$C_1^2 + 2v C_1 \cos \theta - (C^2 - v^2) = 0. \qquad (4)$$

(We must emphasize that equation (4) implies that the three speeds C, $C_1$ and v have been measured with the help of the same clocks, which obviously are clocks not slowed down by motion).

Resolving the second degree equation, we obtain:
$$C_1 = -v \cos \theta \pm \sqrt{C^2 - v^2 \sin^2 \theta}.$$

The condition $C_1 = C$ when $v = 0$ compels us to only retain the + sign so:
$$C_1 = -v \cos \theta + \sqrt{C^2 - v^2 \sin^2 \theta}.$$

- Now, the return of light can be illustrated by the figure 4 below:

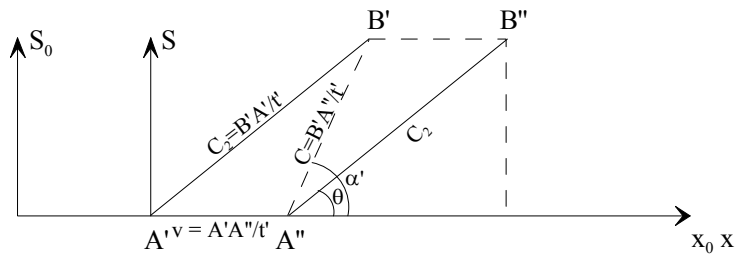

Figure 4

The speed of light is equal to C from B' to A'' and to $C_2$ from B'' to A''.

From the point of view of an observer attached to S, the light comes back to its initial position with the speed $C_2$.

So we can write: $C_2 = \frac{B'A'}{t'}$.

For the observer attached to $S_0$ the light comes from B' to A'' with the speed C, so that

$$C = \frac{B'A''}{t'}.$$

During the light transfer, the system S has moved from A' to A'' with the speed v therefore:

$$v = \frac{A'A''}{t'}.$$

The projection of the speed of light relative to S along the x-axis will be

$$C_2 \cos\theta = C \cos\alpha' + v$$

we easily verify that:

$$(C_2\cos\theta - v)^2 + (C_2\sin\theta)^2 = C^2$$

therefore

$$C_2 = v\cos\theta + \sqrt{C^2 - v^2\sin^2\theta}.$$

The two-way transit time of light along the rod AB, measured with clocks not slowed down by motion, is:

$$2T = \frac{\ell}{C_1} + \frac{\ell}{C_2}. \tag{5}$$

According to the experiment, T must be essentially independent of the angle $\theta$. Therefore, 2T must be equal to:

$$\frac{2\ell_0}{C\sqrt{1 - v^2/C^2}}$$

which is the two way transit time of light along the y direction (previously calculated).

We can see that, in order for this condition to be satisfied, the projection of the rod along the x-axis must shrink in such a way that:

$$\ell\cos\theta = \ell_0\cos\varphi\,\sqrt{1 - v^2/C^2} \qquad \text{(see figure 5)}$$

where $\varphi$ was the angle separating the rod and the $x_0$-axis when the rod was at rest in $S_0$.

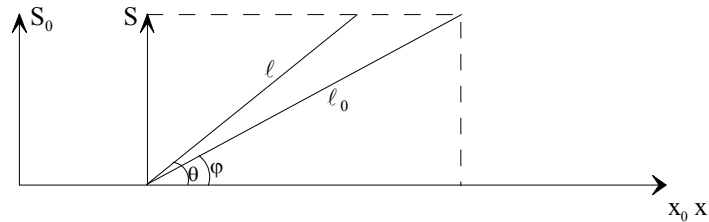

Figure 5
Along the $x_0$, x-axis, the projection of the rod $\ell_0$ contracts,
along the y-axis it is not modified.

from: $\qquad \ell_0\cos\varphi = \dfrac{\ell\cos\theta}{\sqrt{1 - v^2/C^2}}$

and $\qquad \ell_0\sin\varphi = \ell\sin\theta$

we easily verify that: $\qquad \left(\dfrac{\ell\cos\theta}{\sqrt{1 - v^2/C^2}}\right)^2 + (\ell\sin\theta)^2 = \ell_0^{\,2}.$

Finally: $\qquad \ell = \dfrac{\ell_0(1 - v^2/C^2)^{1/2}}{(1 - v^2\sin^2\theta/C^2)^{1/2}}. \tag{6}$

Replacing $\ell$ with this expression in (5) we obtain, as expected:

$$2T = \frac{2\ell_0}{C\sqrt{1-v^2/C^2}}. \tag{7}$$

We conclude that length contraction along the $x_0$, x-axis is a necessary condition so that the two-way transit time of light along a rod given by formula (7) is independent of the orientation of the rod.

- But this is not all. The same conditions combined with clock retardation, enable us to demonstrate that the *apparent* (measured) two way speed of light is C in any direction of space.

Clock retardation is an experimental fact. Let us designate the *apparent* two way transit time of light along the rod in frame S as $2\varepsilon$. We will have (from (7)):

$$\varepsilon = T\sqrt{1-\frac{v^2}{C^2}}$$
$$= \frac{\ell_0}{C}.$$

Now, the length of the rod, measured with a contracted meter stick, is always found equal to $\ell_0$, so that the two-way speed of light is (erroneously) found to be C in any direction of space and independently of the speed v. (As we have seen[26] this is also the case for the *apparent* one-way speed of light measured with the help of clocks synchronized by means of the Einstein-Poincaré synchronization procedure or by slow clock transport). This result is highly meaningful and is a direct consequence of the facts deduced from the Michelson and Morley experiment in vacuo and the experiments and astronomical observations lending support to the anisotropy of the one-way speed of light.

**Note 1**

In our demonstration, although we are indebted to Prokhovnik, we draw different conclusions from his analysis[7]; indeed, since $C = AB'/t$ and $C = B'A''/t'$, it is obvious that t and t' are the real transit times of light along the rod (measured with clocks not slowed down by motion).

Now, since $C_1 = \frac{AB}{t}$ and $C_2 = \frac{B'A'}{t'}$ there is no doubt that $C_1$ and $C_2$ are also measured with the help of clocks not slowed down by motion. This is also the case for $2T = \frac{\ell}{C_1} + \frac{\ell}{C_2}$.

Nevertheless, in his book "The logic of special relativity"[7] chapter " The logic of absolute motion", Prokhovnik identifies the time $2T = \frac{2\ell_0}{C\sqrt{1-v^2/C^2}}$ with the two way transit time of light along the rod, measured with clocks attached to the moving frame.

This cannot be true for the reason indicated above.

(Note that in our notation the moving frame is designated as S, while in Prokhovnik's notation, S designates the aether frame and A the moving frame. We will continue the demonstration with our own notation).

In addition, if Prokhovnik's approach were true, the *apparent* two-way speed of light in frame S would not be C. Indeed, since the standard used for the measurement is also contracted, observer S would find $\ell_0$ for the length of the rod.

Therefore, the *apparent* (measured) two way speed of light in frame S would have been:

$$\frac{2\ell_0}{2\ell_0/C\sqrt{1-v^2/C^2}} = C\sqrt{1-v^2/C^2}$$

which is not in agreement with the experimental facts.

The two-way transit time of light along the moving rod, measured with clocks not slowed by motion is in fact $2\ell_0/C\sqrt{1-v^2/C^2}$, and the *apparent* two-way transit time, measured with clocks attached to frame S, is $2\ell_0/C$. This corresponds to the experimental facts, since, with these values, the *apparent* two-way speed of light in frame S is found equal to

$$2\ell_0 \bigg/ \frac{2\ell_0}{C} = C.$$

Note also that according to aether theory, the real two-way speed of light (measured with non-contracted standards and with clocks not slowed down by motion) can be easily determined from (6) and (7). Along the $x_0$, x-axis we obtain:

$$\frac{2\ell_0 \sqrt{1 - v^2/C^2}}{2\ell_0/C \sqrt{1 - v^2/C^2}} = C(1 - v^2/C^2).$$

As expected, this expression tends to 0 when $v \Rightarrow C$)

**Note 2**

The hypothesis of the aether dragged by the Earth has been generally rejected because of its incompatibility with the theory of aberration. This point of view defended by Lorentz was discussed by Beckmann [31], Mitsopoulos [32] and Makarov [33].

But the theory of the dragged aether is contradicted by the experiment of Lodge [34], who demonstrated that the speed of light is not modified in the neighbourhood of a rotating wheel and by all the experiments and astronomical observations lending support to the anisotropy of the one way speed of light.

## Appendix

In order to justify the lack of experimental evidence, concerning length contraction, Lorentz had recourse to the law of variation of mass with speed $m = m_0 \gamma$. Yet in the light of our present day knowledge, it is necessary to verify if the Lorentz aether is compatible with mass variation.

Indeed, in another publication[35], we have demonstrated that, if we assume the Lorentz aether, the experimental space-time transformations can be derived from the Galilean transformations by subjecting them to the three kinds of distortions brought about by length contraction, clock retardation and clock synchronization with light signals.
Therefore the experimental space-time transformations conceal hidden variables that are nothing else than the Galilean transformations which are the true transformations (not altered by measurement distortions).

(Of course this implies that when a body A moves at speed $v_A$ from the origin of a co-ordinate system which is at rest with respect to the aether frame, the speed relative to A of another body B moving along the direction 0A will be limited to:

$v_B < c - v_A$)

At first sight these transformations do not seem compatible with the law of variation of mass with speed, since in order to demonstrate this law we generally make use of Einstein's relativity principle which assumes the law of conservation of the relativistic momentum in any inertial frames and does not assume the Galilean transformations. As a consequence, the other Lorentz assumptions do not seem in accordance with the law $m = m_0 \gamma$ obtained by experiment. For this reason these Lorentz assumptions appeared questionable to us in an earlier publication.

Yet the objection can be challenged. To this end we shall resort to the following demonstration:
Consider a body *at rest in the fundamental frame*, which is subjected to a force *F*. The elementary expression for the kinetic energy acquired by the body in the displacement $d\ell$ is:

$dE_C = Fd\ell$,

where $Fd\ell$ is the work carried out by the force *F* in this displacement. (We suppose that *F* and $d\ell$ are aligned).
Now, the equivalence of mass and energy takes the form
$E = mC^2 = E_C + m_0 C^2$, (8)
As pointed out by Rohrlich[36] mass-energy equivalence can be demonstrated without the help of the Lorentz transformations.
From (8) we can write:
$dE = Fd\ell$ . (9)

From (8) and (9) we obtain successively:

$$C^2 dm = \frac{d(mv)}{dt} v dt,$$

$$= (m\frac{dv}{dt} + v\frac{dm}{dt})v dt,$$

$$C^2 dm = mv dv + v^2 dm,$$

$$\frac{dm}{m} = \frac{v dv}{C^2 - v^2}.$$

Designating $C^2 - v^2$ as $u$ so that $v\, d\, v = -du/2$, we then find

$$Log\, m = -\frac{1}{2} Log(C^2 - v^2) + Log\, k$$

$$= Log\, k(C^2 - v^2)^{-1/2}$$

and

$$m = \frac{k}{C\sqrt{1 - v^2/C^2}}.$$

For $v = 0 \Rightarrow m = k/C = m_0$, thus:

$$m = \frac{m_0}{\sqrt{1 - v^2/C^2}}, \tag{10}$$

where $m_0$ is the rest mass.
(See also the alternative demonstration given by Selleri[37] on the basis of a work by Lewis[38]).
Expression (10) is completely exact only when $m_0$ is the mass at rest in the fundamental frame.
Indeed between two 'inertial' systems $S_1$ and $S_2$ associated to bodies moving at speeds $v_{01}$ and $v_{02}$ relative to the aether system $S_0$ the expressions

$$m_1 = \frac{m_0}{\sqrt{1 - v_{01}^2/C^2}} \quad \text{and} \quad m_2 = \frac{m_0}{\sqrt{1 - v_{02}^2/C^2}}$$

yield

$$m_2 = m_1 \frac{\sqrt{1 - v_{01}^2/C^2}}{\sqrt{1 - v_{02}^2/C^2}},$$

which to first order gives:

$$m_2 \approx m_1[1 + \frac{1}{2}v_{12}^2/C^2 + v_{01}v_{12}/C^2].$$

This expression is different and obviously greater than the relativistic expression. Indeed since according to relativity, $m_0$ is the rest mass in all inertial frames, we have:

$$m_2 = \frac{m_0}{\sqrt{1 - v_{12}^2/C^2}} \approx m_0(1 + \frac{1}{2}v_{12}^2/C^2).$$

In the same way, from $\ell_2 = \ell_0\sqrt{1 - v_{02}^2/C^2}$ and $\ell_1 = \ell_0\sqrt{1 - v_{01}^2/C^2}$
we infer:

$$\ell_2 = \ell_1 \frac{\sqrt{1 - v_{02}^2/C^2}}{\sqrt{1 - v_{01}^2/C^2}}.$$

Therefore the relativity principle, does not apply to real values of the measurements, but we have shown in ref[35] that with the usual measurements which are performed with contracted meter sticks and retarded clocks, synchronized with light signals, the experimental space-time transformations assume a mathematical form identical to the Lorentz transformations, (although their meaning is quite different) and, therefore, with these transformations, the (*apparent*) laws of physics including $m = m_0\gamma$ and $\ell = \ell_0/\gamma$ assume an identical mathematical form *in any 'inertial' frame.*

*This argument, which enables to surmount the objections raised against the Lorentz approach, merely confirms,(in agreement with the other arguments developed in the text), the coexistence of the Lorentz assumptions and the experimental (apparent) law of mass increase, despite what differentiates them.*